\documentclass[11pt,letterpaper]{JHEP3}
\usepackage[dvips]{epsfig}
\usepackage{amsmath,amssymb,epsf,amsfonts}
\usepackage{graphicx}
\usepackage{dcolumn}
\usepackage{bm}
\title{Chiral anomalies and AdS/CMT in two dimensions}
\author{ Kristan Jensen
\vspace{0.4cm} \\
\small  Department of Physics, University of Victoria,
Victoria, BC V8W 3P6, Canada\\
\small Department of Physics, University of Washington, Seattle, WA 98195-1560 , USA \\
\small  ~\,E-mail: \tt{kristanj@uvic.edu}
}

\maketitle

\abstract{I clarify some recent confusion regarding the holographic description of finite-density systems in two dimensions.  Notably, the chiral anomaly for symmetry currents in 2d conformal field theories (CFT) completely determines their correlators.  The important exception is a CFT with a gauge theory to which we may couple an external current, as in the probe D3/D3 system or the putative dual to the charged BTZ black hole.  These systems are analyzed with an eye for potential condensed matter applications.}

\begin{document}

\section{Introduction}
One of the most beautiful examples of holography is the duality between Chern-Simons (CS) theory on a three-manifold with boundary and a chiral Wess-Zumino-(Novikov)-Witten (WZW) model on the boundary~\cite{Witten:1988hf}.  The latter is a conformal field theory (CFT), and so this is a duality between a bulk theory and a boundary CFT.  The WZW model furnishes a representation of a chiral current algebra on the boundary~\cite{Witten:1983ar} so that the bulk gauge field is related to the boundary symmetry current .  On the other hand, the AdS/CFT correspondence~\cite{Maldacena:1997re,Witten:1998qj,Gubser:1998bc} is a wide class of dualities between gravitational theories on AdS$_{d+1}$ spacetimes and CFTs living on the $d$-dimensional AdS boundary.  The correspondence also relates bulk gauge fields to boundary symmetries.  How do these pictures intersect?  That is, how does AdS/CFT work for the special case of a 2d boundary CFT with some symmetries (which are enlarged by the conformal symmetry to current algebras)?

This question was rigorously answered in~\cite{Gukov:2004id}: in stringy examples of AdS/CFT, the gravitational theory for the AdS$_3$ gauge field typically includes both Chern-Simons and Maxwell terms.  That is, there is a massive gauge field in the bulk dual to a higher-dimension vector primary operator in the boundary theory.  Meanwhile the flat part of the gauge field is dual to a chiral symmetry current through the Chern-Simons term, precisely as in the simpler Chern-Simons/WZW duality.  Indeed the CS/WZW duality largely survives.

The holography of two-dimensional CFTs is then rather special.  In all higher-dimensional examples, the propagating modes of a bulk gauge field are dual to a symmetry current in the boundary CFT.  Here the currents are captured by topological terms in the bulk.  This fact has stark consequences for the so-called AdS/CMT program, that is the use of holography to realize and study condensed matter phenomena.  Most of the AdS/CMT applications study features of the boundary theory at nonzero charge density, notably including (i.) transport~\cite{Son:2006em} and (ii.) spontaneous symmetry breaking in holographic superfluids~\cite{Hartnoll:2008vx}.  As we will see, the bulk Chern-Simons terms imply that 2d CFTs at nonzero density behave rather differently.

Indeed, there have been a number of recent papers that holographically study 2d theories at nonzero density~\cite{Hung:2009qk,Colgain:2010rg,Balasubramanian:2010sc}.  A number of these works~\cite{Maity:2009zz,Ren:2010ha} are phenomenological and, mirroring higher-dimensional AdS/CMT studies, do not include Chern-Simons terms.  As such they do \emph{not} study the physics of 2d CFTs with current algebras.  Nonetheless, the rules of AdS/CFT can be employed to constructively \emph{define} a field theory dual to these systems.  We are left with at least two questions.  First, what \emph{are} the physics of a 2d CFT at nonzero density?  Second, how should we think of the setup with no Chern-Simons term?

The purpose of this letter is answer these questions.  First, in Section~\ref{holoCS} I review 3d Chern-Simons theories and how they arise in holography.  I use these results to compute connected correlators involving the current.  Remarkably, these correlators are completely determined by the bulk Chern-Simons term.  As a result the currents never behave hydrodynamically.  In Section~\ref{anomalies} I find a purely field-theoretic explanation for the result.  It essentially follows from Schwinger's solution of two-dimensional QED.  Next, I clarify the dual of Einstein-Maxwell theory on AdS$_3$ in Section~\ref{chargedBTZ}: the bulk gauge theory is dual to a weakly coupled gauge theory on the boundary to which we may couple an external current.  This result was anticipated in~\cite{Marolf:2006nd}.   Such a system can be embedded in string theory through the D3/D3 probe brane system.  Employing vector/scalar duality in the bulk, I holographically renormalize Einstein-Maxwell theory on AdS$_3$\footnote{A number of works have studied just this problem, proposing various regularization schemes.} and use this result to compute a number of observables in the dual theory.  The two most interesting results there are (i.) a Weyl anomaly of the form $T_a^a\sim \text{tr}(j^2)$ and (ii.) a proper renormalization of the charged BTZ black hole~\cite{Banados:1992wn,Martinez:1999qi} from the point of view of the boundary CFT.  I conclude with some thoughts on the prospects of AdS/CMT applications for 2d theories.

\section{Holography and Chern-Simons theory}
\label{holoCS}
Let us begin with a review of current algebras in 2d CFTs and how they appear in AdS/CFT.  From here we will be positioned to study their correlators.

\subsection{Some review}
Consider a CFT on a Riemann surface $M$ without boundary.  Suppose that the theory has some chiral symmetry currents.  Left and right-moving currents are really holomorphic and anti-holomorphic in two dimensions.  Conformal invariance demands that the operator product expansion (OPE) of two currents is
\begin{align}
\label{OPE}
j^a(z)j^b(0)=-\frac{kd^{ab}}{4\pi^2z^2}+\frac{f^{abc}j^c(0)}{2\pi z}+\text{non-singular},
\end{align}
where $z$ and $\bar{z}$ are holomorphic and anti-holomorphic coordinates on $M$, $d$ is the group metric, and the $f^{abc}$ are the structure constants of the algebra of the symmetry group $G$.  The anti-holomorphic currents satisfy a similar relation.  The OPE then implies that the symmetry algebra, like the conformal algebra, is enlarged by conformal symmetry to be infinite dimensional.  To see this decompose the currents into the modes
\begin{equation}
j^a_z(z) = \sum_n\frac{j^a_n}{z^{n+1}}, \hspace{.5in} j^b_{\bar{z}}(\bar{z})=\sum_m\frac{\bar{j}^b_m}{\bar{z}^{m+1}}.
\end{equation}
These generate position-dependent symmetry transformations.  The OPE Eq.~(\ref{OPE}) implies that the modes satisfy
\begin{equation}
\label{kacMoody}
[j_n^a,j_b^m]=-\frac{nkd^{ab}}{4\pi^2}\delta_{n+m,0}+\frac{f^{abc}}{2\pi}j^c_{n+m},
\end{equation}
For $n=m=0$ this is the ordinary symmetry algebra.  It is enlarged to the infinite-dimensional Eq.~(\ref{kacMoody}), which is called an affine Lie algebra or Kac-Moody algebra.  The representations of affine Lie algebras are classified by the ``level'' $k$: there are a finite number of irreducible representations at level $k$, all of which may be constructed from tensor products of a unique representation for $k=1$.  The level is itself quantized to be a positive integer.

Chiral currents are anomalous.  This fact is encoded in the first term in Eq.~(\ref{OPE}).  To see this, consider turning on a source, $A^a_{\bar{z}}(z,\bar{z})$, for the holomorphic currents.  Then we may obtain the one-point function of $j$ in an expansion in powers of $A$ whose first term is
\begin{equation}
\langle j_z^a(z)\rangle = \int d^2z' \langle j_z^a(z)j_z^b(z')\rangle A^b_{\bar{z}}(z',\bar{z'})+\hdots
\end{equation}
where the correlators are computed at $A=0$.  Taking the divergence of both sides and using $\partial_{\bar{z}}\frac{1}{\bar{z}^2}=-2\pi\partial_{z}\delta^{(2)}(z,\bar{z})$ we find an anomaly
\begin{equation}
\label{chiAnom}
D_{\bar{z}}j^a_z(z)=\frac{kd^{ab}}{2\pi} \partial_z A^b_{\bar{z}}(z,\bar{z}),
\end{equation}
which is the usual 2d chiral anomaly.  This is a non-perturbative result: the only nonzero contribution comes from the $1/z^2$ term in the $jj$ OPE.  The level then directly corresponds to the strength of the anomaly.

As in higher dimensions we may write down an effective theory for the currents~\cite{D'Adda:1982es}.  At its fixed points such a theory gives a Lagrangian description of the algebra Eq.~(\ref{kacMoody}).  In two dimensions this is
\begin{equation}
\label{chiralWZW}
S_{\rm eff}=\frac{1}{4 \lambda^2}\int_{M} d^2z\,\text{tr}( g^{-1}\partial_z g g^{-1}\partial_{\bar{z}}g)+\frac{k}{12\pi}\int_N\text{tr}(g^{-1}dg)^3
\end{equation}
where the second piece is the Wess-Zumino (WZ) term~\cite{Wess:1971yu} which encodes the anomaly.  The theory lives on the spacetime $M$ and involves a $G$-valued field $g$ on $M$.  $M$ may then be thought of as a two-surface inside $G$.  Since $\pi_2(G)=0$, the map $g$ may be extended to a solid three-surface $N\subset G$ whose boundary is $M$.  The WZ term has two important properties.  First, its integrand is closed, 
\begin{equation}
d(g^{-1}dg)^3=0.
\end{equation}
As a result the WZ term may be represented locally as the integral of a two-dimensional action density on $M$; Eq.~(\ref{chiralWZW}) indeed describes a 2d theory.  Second, the WZ term is only defined modulo $2\pi k$ (for $g$ in the fundamental representation and $G$ simply connected).  This ambiguity is related to the fact that the extension of $M$ to $N$ is not unique: there are $\pi_3(G)=\mathbb{Z}$ topologically inequivalent ways to extend $g$ into a map from $N$ to $G$.  Consider two inequivalent extensions $N$ and $N'$.  Glue them along their common boundary, $M$, to give a three-surface $\tilde{N}$ without boundary.  Integrating the WZ term over $\tilde{N}$ indeed yields $2\pi k$ times an integer: the integer is just the winding number of the map.  Thus the WZ term integrated over $N$ is only defined modulo $2\pi k$ as claimed.  As a result the weight $\exp(iS)$ is well-defined for integer $k$ and so Eq.~(\ref{chiralWZW}) gives a perfectly well-behaved quantum field theory.

As Witten showed long ago~\cite{Witten:1983ar} Eq.~(\ref{chiralWZW}) describes an asymptotically free sigma model with an infrared fixed point at $\lambda^2=\pi/|k|$.  Indeed, for positive $k$ the current $j_z=\frac{k}{2\pi} g^{-1}\partial_{\bar{z}}g$ satisfies the algebra Eq.~(\ref{kacMoody}) at level $k$ for this value of $\lambda$.  Conversely, for negative $k$ the current $j_{\bar{z}}=-\frac{k}{2\pi} g^{-1}\partial_{\bar{z}}g$ satisfies an affine Lie algebra at level $-k$.  At the fixed point the theory is called a chiral Wess-Zumino-(Novikov)-Witten model.

\subsubsection{Holography}
The canonical example of the AdS/CFT correspondence is the duality between type IIB string theory on the background AdS$_5\times\mathbb{S}^5$ and $\mathcal{N}=4$ $SU(N)$ super Yang-Mills (SYM) theory in $(3+1)$-dimensions~\cite{Maldacena:1997re}.  The field theory is conformal and has two dimensionless parameters: $N$ and the 't Hooft coupling $\lambda = g_{\rm YM}^2 N$.  Meanwhile, the 10d geometry is supported by $N$ units of five-form Ramond-Ramond (RR) flux through the AdS$_5$ and $\mathbb{S}^5$; each space has a radius $l$ that is related to field theory quantities by $l^4=4\pi \lambda \alpha'^2$.  That is the stringy picture becomes weakly curved supergravity in the double scaling limit $\lambda,N\rightarrow\infty, \frac{\lambda}{N}\rightarrow 0$.  This is a common result in holography: the gravitational dual is tractable when the field theory is strongly coupled and vice versa.

The dictionary for the duality is that both theories have the same generating functional~\cite{Witten:1998qj,Gubser:1998bc} so that in the double scaling limit,
\begin{equation}
\label{generatingFunc}
\langle \exp\left[  \int d^4x \sqrt{-g}J(x)\mathcal{O}(x)\right] \rangle_{\rm CFT}=e^{S_{\rm bulk}[\phi]},
\end{equation}
where $J$ is the source for a dimension $\Delta$ operator $\mathcal{O}$ which is dual to a bulk field $\phi$.  Similar terms may be added for symmetry currents (dual to bulk gauge fields) and the stress tensor (dual to the bulk metric). In order for Eq.~(\ref{generatingFunc}) to make sense, we must relate the source $J$ to the bulk field $\phi$.  We usually do this by identifying the $J$ and leading term in the near-boundary falloff of $\phi$.  With this identification we may compute correlators of $\mathcal{O}$ by taking derivatives of both sides with respect to $\phi^{(0)}$.  The conformal invariance of the LHS (in the vacuum) is inherited from the $SO(4,2)$ isometry of the dual AdS$_5$ geometry.

Of course there is nothing in Eq.~(\ref{generatingFunc}) that specifically refers to $\mathcal{N}=4$ SYM or string theory.  Eq.~(\ref{generatingFunc}) may be viewed as a recipe for defining a dual field theory, provided that the bulk theory is consistent.  In this way any stable string theory vacuum with an AdS factor is dual to a CFT.  

\subsubsection{Chern-Simons from compactification}
\label{sugra}
It is well-known that Chern-Simons terms naturally arise in AdS$_3$ compactifications of supergravity.  I will illustrate this fact with an example: the $U(1)$ gauge theory in the D1/D5 system~\cite{Gukov:2004ym}.  We begin with $N_1$ D1 branes (extended along the $01$ directions) and $N_5$ D5 branes (extended along $016789$) wrapping a compact Calabi-Yau two-fold $M_4$ in the $6789$ directions, i.e. either $T^4$ or $K3$.  Such a setup preserves eight supercharges of supersymmetry.  In the usual holographic limits, the near-horizon geometry is AdS$_3\times\mathbb{S}^3\times M_4$.  That is, the metric and dilaton in the near-horizon region are
\begin{equation}
g=(r_1r_5)g_{\text{AdS}_3}+(r_1r_5) d\Omega^2_3+\frac{r_1}{r_5}g_{M_4},\hspace{.5in} e^{\phi}=g_s \frac{r_1}{r_5}
\end{equation}
where the AdS$_3$ and $\mathbb{S}^3$ have radius $l^2=r_1r_5$ and the $r_i$ are
\begin{equation}
r_1^2=g_s \alpha' \frac{(2\pi\sqrt{\alpha'})^4}{V_4}N_1, \hspace{.5in} r_5^2=g_s\alpha' N_5,
\end{equation}
where the volume of $M_4$ is $V_4$.  The geometry is supported by the RR three-form flux
\begin{equation}
g_sF_3= \frac{2}{r_1^2} \omega_0+2r_5^2 \omega_1,
\end{equation}
where $\omega_0$ and $\omega_1$ are the volume forms on unit AdS$_3$ and $\mathbb{S}^3$ respectively.  The dual field theory is a little complicated; it can be described by a sigma model with target space being the moduli space of instantons on $M_4$.  See the seminal~\cite{Seiberg:1999xz} for more information.

In the $T^4$ compactification there are a number of AdS$_3$ $U(1)$ gauge fields.  The easiest ones to find are those from the metric and RR three-form.  Following~\cite{Gukov:2004ym}, consider a modified ansatz for the geometry,
\begin{align}
g&=l^2 g_{\text{AdS}_3}+l^2 d\Omega_3^2 +\frac{r_1}{r_5}  \sum_{i=1}^4L_i^2 (d\theta_i+a_i)^2, \\
\nonumber
g_sF_3&= \frac{2}{r_1^2} \omega_0+2r_5^2 \omega_1+\sum_{i=1}^4 db_i\wedge d\theta_i,
\end{align}
where the $\theta_i$'s have domain $[0,1)$ and the $a_i$ and $b_i$ are AdS$_3$ gauge fields.  The volume of the $T^4$ is just $V_4=\prod_i L_i$.  The relevant part of the IIB supergravity action is
\begin{equation}
S=\frac{1}{2\kappa_{10}^2}\int d^{10}x \sqrt{-g}e^{-2\phi}R-\frac{1}{4\kappa_{10}^2}\int F_3\wedge \star F_3+\hdots
\end{equation}
where $2\kappa_{10}^2=g_s^2(2\pi)^3(2\pi\alpha')^4$ is the 10d gravitational constant.  The kinetic term for $F_3$ contains the quadratic term for the $b_i$ \emph{as well as} a Chern-Simons term.  Together with the kinetic term for the $a_i$ the effective AdS$_3$ action is easily computed to be
\begin{equation}
S_3=\frac{N_1}{e^2}\int_{\text{AdS}_3}\left(- \frac{L_i^2l}{2}\,da_i\wedge \star da_i-\frac{l}{2L_i^2}\,db_i\wedge \star db_i+2db_i\wedge a_i\right),
\end{equation}
where $e^2=4\pi g_s^3 \alpha'$ is found after straightforward dimensional reduction.  By the useful change of variable
\begin{equation}
A_i^{(\pm)}\equiv \frac{1}{\sqrt{2}}\left( \frac{1}{L_i}b_i\mp L_i a_i\right),
\end{equation}
the effective action becomes
\begin{equation}
\label{paritySeff}
S_3=\frac{N_1}{e^2}\int \left(-\frac{l}{2}dA_i^{(+)}\wedge \star dA_i^{(+)}-A_i^{(+)}\wedge dA_i^{(+)}\right)+\left( -\frac{l}{2}dA_i^{(-)}\wedge\star dA_i^{(-)}+A_i^{(-)}\wedge dA_i^{(-)}\right).
\end{equation}
Eq.~(\ref{paritySeff}) describes a parity-invariant $(U(1)\times U(1))^4$ Chern-Simons-Maxwell (CSM) theory (bulk parity is the combined operation $x,A_i^{(\pm)}\rightarrow -x,A_i^{(\mp)}$).  There is also another parity-invariant $(U(1)\times U(1))^4$ CSM theory from the RR five-form and NS three-form.  The CS term for that theory descends from the 10d CS term.

The equations of motion for $A_i^{(\pm)}$ are those of $m^2l^2=4$ gauge fields in AdS$_3$.  By Eq.~(41) of Ref.~\cite{Larsen:1998xm} the $A_i^{(\pm)}$ are then dual to dimension $(2,1)$ and $(1,2)$ vector primary operators.  Meanwhile, as we are about to see in great detail, the flat parts of $A_i^{(\pm)}$ are dual to $(1,0)$ and $(0,1)$ $U(1)$ current algebras at level $N_1$.  Finally, there is also an $SU(2)\times SU(2)$ gauge sector that descends from the $SO(4)$ isometry of the $\mathbb{S}^3$.  Those gauge fields may also be arranged into a parity-even CSM theory but the derivation is a little more subtle~\cite{Hansen:2006wu}.

\subsection{Currents from Chern-Simons holography}
Having observed that Chern-Simons terms are natural in AdS$_3$ compactifications, what does this imply for the physics of the dual field theory?  To answer this question, I will consider a more general bulk action for a non-Abelian gauge field $A$ with gauge group $G$,
\begin{equation}
\label{csSbulk}
S_{\rm bulk}=-\frac{k}{4\pi}\int \text{tr}\left( A\wedge dA+\frac{2}{3}A\wedge A\wedge A\right)+\hdots
\end{equation}
where the dots indicate arbitrary Maxwell and higher-derivative corrections and $k$ is integer.  For an asymptotically AdS$_3$ space,
\begin{equation}
G=\frac{r^2}{l^2}(g^{(0)}(x)+O(r^{-2})) +\frac{l^2}{r^2}dr^2,
\end{equation}
$A$ falls off near the boundary $r\rightarrow\infty$ as
\begin{equation}
A=A^{(0)}+r^{1-\Delta}A^{(2)}+\hdots
\end{equation}
where the propagating part of the gauge field is dual to a dimension $\Delta$ vector primary and $A^{(0)}$ is flat.  Also, there is a conformal class of metrics induced on the boundary of which $g^{(0)}$ is a representative.

There are several observations in order.  First, the bulk CS term is equivalent to the presence of an anomaly in the dual theory~\cite{Witten:1998qj}.  The CS term is therefore \emph{required} to describe a dual symmetry current.  The second observation is that the bulk theory is not well-defined on a manifold $M$ with boundary $\partial M$: its action has no extrema, as may be seen by varying $A\rightarrow A+\delta A$,
\begin{equation}
\label{CSvary}
\delta S_{CS} = -\frac{k}{2\pi}\int_M \text{tr}(\delta A\wedge (dA+A\wedge A)-\frac{k}{4\pi}\int_{\partial M} \text{tr}(\delta A \wedge A).
\end{equation}
This fact has two consequences: (i.) the boundary term implies that $A^{(0)}$ are \emph{both} positions and momenta and (ii.) the CS term must be supplemented with a boundary term and appropriate boundary conditions on $A$.  The form of the boundary term will depend on the boundary conditions we impose.  A typical way to separate $A$ into positions and momenta is to use the complex structure on the boundary.  For $k$ positive, a good boundary condition is to fix $A^{(0)}_{\bar{z}}$ while for $k$ negative we fix $A^{(0)}_z$.  The right boundary term for either case is then
\begin{equation}
\label{csCT}
S_{\rm CT}=\frac{|k|}{4\pi}\int_{\partial \text{AdS}_3} d^2z\, \text{tr}(A_z A_{\bar{z}}).
\end{equation}
While the CS term does not depend upon the bulk metric the boundary term does through the complex structure~\cite{Kraus:2006wn}.

The third observation is that the gravitational action is not quite gauge-invariant.  It transforms by a boundary term under $A\rightarrow g^{-1}(d+A)g$ which for $k$ positive is
\begin{equation}
\Delta(S_{\rm bulk}+S_{\rm CT})=\frac{k}{4\pi}\int_{\partial \text{AdS}_3} d^2z\,\text{tr}\left( g^{-1}\partial_z g g^{-1}\partial_{\bar{z}}g+2 g^{-1}\partial_z g A^{(0)}_{\bar{z}}\right) +\frac{k}{12\pi}\int_{\text{AdS}_3}\,\text{tr}(g^{-1}dg)^3.
\end{equation}
This is just the chiral WZW action Eq.~(\ref{chiralWZW}) with a coupling of $j_z=\frac{k}{2\pi}g^{-1}\partial_z g$ to an external gauge field $A^{(0)}_{\bar{z}}$.  The bulk path integral over $A$ therefore includes a chiral WZW model at level $k$ which corresponds to the $G$ chiral symmetry of the dual theory.  This is the classic duality between WZW models and CS theory~\cite{Witten:1988hf,Elitzur:1989nr}.

Holography gives us another means to compute correlators of the current.  Varying $A\rightarrow A+\delta A$ with $\delta A = \delta A^{(0)}_zdz+\delta A^{(0)}_{\bar{z}}d\bar{z}+\hdots$ gives
\begin{equation}
\delta \left(S_{\rm bulk}+S_{\rm CT}\right) = \frac{k}{2\pi}\int d^2z \,\text{tr}(\delta A^{(0)}_{\bar{z}}A^{(0)}_z).
\end{equation}
The one-point function of the dual current is
\begin{equation}
\label{onePoint}
\langle j^a_z\rangle =\frac{g^{(0)}_{z\bar{z}}}{\sqrt{g^{(0)}}}\frac{\delta S_{\rm grav}}{\delta A^{(0)a}_{\bar{z}}}= \frac{k}{2\pi}d^{ab}A^{(0)b}_z,
\end{equation}
Combined with the flatness of $A^{(0)}$, we find the chiral anomaly Eq.~(\ref{chiAnom}).  Similar results hold when we consider negative $k$.  The flat $A^{(0)}$ is then dual to an anti-holomorphic current $j_{\bar{z}}$ at level $-k$.

As I mentioned above, the boundary term Eq.~(\ref{csCT}) depends on the boundary metric.  Writing the boundary term as
\begin{equation}
S_{\rm CT} = \frac{|k|}{8\pi}\int_{\partial \text{AdS}_3} d^2x \sqrt{g^{(0)}}\,\text{tr}(A^2),
\end{equation}
we see that the flat part of the gauge field contributes to the stress tensor through this term.  In fact this is its only contribution to the stress tensor.  The current algebra part of the boundary stress tensor is then
\begin{equation}
T_{ab}=\frac{|k|}{4\pi}\text{tr}\left( A_aA_b-\frac{1}{2}g^{(0)}_{ab} A^2\right),
\end{equation}
where $a,b$ index boundary coordinates.  Decomposing $T$ into holomorphic and anti-holomorphic parts and employing Eq.~(\ref{onePoint}) for $k$ positive we get
\begin{equation}
T_{zz}=\frac{\pi}{k}\text{tr}(j_z j_z), \hspace{.3in} T_{\bar{z}\bar{z}}=\frac{k}{4\pi}\text{tr}\left( ^{(0)}_{\bar{z}}A^{(0)}_{\bar{z}}\right), \hspace{.3in} T_{z\bar{z}}=0.
\end{equation}
This is just the stress tensor for a WZW model.  This is important: the stress tensor for the current algebra theory decouples from the rest of the stress tensor (dual to the AdS$_3$ graviton) and is determined by the boundary term.  This is the same term as the one that describes the boundary stress tensor in CS/WZW holography.  The CS/WZW duality therefore survives largely intact (modulo subtle details; see~\cite{Gukov:2004id}) for AdS/CFT holography.

\subsubsection{Results for abelian theories}
\label{abelian}
I will now specialize to the case $G=U(1)$, answering several questions related to potential AdS/CMT applications.  First, what is a(n equilibrium) charged black hole for this theory?  This question was actually answered some time ago~\cite{Kraus:2005vz}.  Suppose that the bulk geometry has a horizon, e.g. the BTZ black hole.  Then $A$ must have zero holonomy around the Euclideanized time circle at the horizon.  If the system is translationally invariant in the field theory directions, this is just the boundary condition $A_0(r=r_h)=0$.  Moreover, if we do not excite the higher dimension vector primary the bulk gauge field will be flat everywhere.  Then the spatial component of $A^{(0)}$, $A^{(0)}_1$, is a free parameter that indexes a one-parameter family of black holes supplemented with either a holomorphic or anti-holomorphic current density.  For $k$ positive I define the anti-holomorphic chemical potential $A^{(0)}_{\bar{z}}\equiv \mu_{\bar{z}}$ and so by Eq.~(\ref{onePoint}) get $\langle j_z\rangle=\frac{k}{4\pi}\mu_{\bar{z}}$.

There is also a parity-invariant charged black hole with nonzero charge but zero current if the bulk CS theory is parity-invariant.  That is if the bulk theory is of the form
\begin{equation}
S_{\rm bulk}\sim -\frac{k}{8\pi}\int (A_1\wedge dA_1-A_2\wedge dA_2)+\hdots
\end{equation}
The dual theory has a holomorphic and an anti-holomorphic $U(1)$ current algebra, both at level $k$.  The bulk theory is invariant under the combined operation $x,k\rightarrow -x,-k$, which is defined to be parity.  It exchanges the left and right-moving currents while flipping space.  Then there is a two-parameter family of charged black holes with
\begin{equation}
\langle j_z\rangle = \frac{k}{4\pi}\mu_{\bar{z}}, \hspace{.5in} \langle j_{\bar{z}}\rangle=\frac{k}{4\pi}\mu_z.
\end{equation}
These may be combined into the vector and axial currents $j\equiv j_z dz+j_{\bar{z}}d\bar{z}$, $j_A\equiv j_z dz-j_{\bar{z}}d\bar{z}$.  The black holes with all (vector) charge and no (vector) current are then those with $\mu_z=\mu_{\bar{z}}$.

What about a completely general charged black hole?  That is, consider a geometry with a horizon and any number of (uncharged) bulk fields turned on with appropriate boundary conditions.  For the gauge field this will mean that $A_0(r_h)=0$ and that the anti-holomorphic (holomorphic) part of $A$ is fixed at the boundary.  The geometry will be dual to the field theory in some ensemble with some control parameters.  We then immediately know the bulk dual for the same theory with a shifted chemical potential: simply shift the bulk gauge field by an arbitrary $A_1(x^1)$.  For $k$ positive the current and chemical potential are shifted as $ \langle \Delta j_z(x)\rangle = \frac{k}{4\pi}\Delta \mu_{\bar{z}}(x)$.  I stress the condition that the charged fields in the bulk have trivial profiles.  When this is not true, the equations of motion for the charged fields will no longer be satisfied after shifting $A_1$.

Let us take this one step further.  Suppose that the boundary is just flat (Euclidean) space.  Then an arbitrary solution to the bulk equations of motion (again, provided that only uncharged fields have nonzero profiles) may be amended by a flat shift of $A$,
\begin{equation}
\delta A_{\mu}dx^{\mu}=c q_a e^{i q_a x^a}dx^a,
\end{equation}
where $\mu$ labels bulk coordinates and $a=0,1$ labels boundary coordinates.  This solution is consistent with the infrared boundary condition and shifts the boundary current by
\begin{equation}
\label{euc2pt}
\langle \delta j_{z}(q)\rangle = \frac{k}{4\pi}\frac{(q_0-iq_1)^2}{q^2}\delta \mu_{\bar{z}}(-q),
\end{equation}  
which gives the two-point function of the current
\begin{equation}
\langle j_z(q) j_z(-q)\rangle = \frac{k}{4\pi}\frac{(q_0-iq_1)^2}{q^2}.
\end{equation}
This may be Fourier transformed to real space giving $\langle j_z(z)j_z(0)\rangle = -\frac{k}{8\pi^2}\frac{1}{z^2}$, precisely matching the OPE Eq.~(\ref{OPE}).

Suppose that we have a parity-even combination of currents.  Then combining the two-point function of $j_z$ with that of $j_{\bar{z}}$ we have the correlator for the vector current
\begin{equation}
\label{jjEuc}
\langle j^{a}(q)j^b(-q)\rangle = -\frac{k}{2\pi}\frac{2\epsilon^{ac}q_c\epsilon^{bd}q_d-q^2\delta^{ab}}{q^2},
\end{equation}
with $\epsilon^{01}=1$.  The first term corresponds to an anomaly for the axial current $j_A^a=\epsilon^{ab}j_b$ in the presence of a vector field strength and the second to an anomaly for $j$ in an axial field strength.  As is well-known, we may add a local counterterm to the theory to eliminate the second term.  The term is proportional to $\int d^2z\, \text{tr}(A_{1\bar{z}}A_{2z})$, where $A_1$ and $A_2$ couple to $j_z$ and $j_{\bar{z}}$ respectively.  The two-point function in this gauge-invariant theory is just the first term of Eq.~(\ref{jjEuc}),
\begin{equation}
\label{gaugeInvJJ}
C^{ab}(q)\equiv\langle j^{a}(q)j^b(-q)\rangle = -\frac{k}{\pi}\frac{\epsilon^{ac}q_c\epsilon^{bd}q_d}{q^2}.
\end{equation}
We may analytically continuing to real-time and find,
\begin{equation}
C^{00}(\omega,q)=\frac{k}{\pi}\frac{q^2}{\omega^2-q^2}, \hspace{.3in} C^{01}(q)=\frac{k}{\pi}\frac{\omega q}{\omega^2-q^2}, \hspace{.3in} C^{11}(q)=\frac{k}{\pi}\frac{\omega^2}{\omega^2-q^2}.
\end{equation}
Then the current (subtracting off the potential piece at zero chemical potential) satisfies Ohm's law $j_1(\omega)=\sigma(\omega)F_{01}(\omega)$ with real and imaginary conductivities 
\begin{equation}
\sigma(\omega)=k \delta(\omega)+ \frac{ik}{\pi\omega}.
\end{equation}
Moreover, the absence of any higher $n$-point connected correlators of the current implies that Ohm's law receives no higher-order corrections in $F_{01}$.  There is no non-linear transport.

This is a remarkable set of statements.  The $U(1)$ currents are essentially non-interacting for \emph{any} (renormalizable) deformation of the boundary CFT\footnote{This result was already noted for the current-current correlator of a 2d CFT at nonzero temperature in~\cite{Kovtun:2008kx}.  That result is guaranteed by conformal invariance.}.  Their correlators are completely determined by the Chern-Simons part of the bulk action and so by the anomaly.  The one caveat is that this result only holds if charged fields in the bulk have trivial profiles.

\paragraph{Aside: probe branes}
\label{8NDprobe}
Chern-Simons theories are also found in a number of probe brane systems~\cite{Karch:2002sh}.  These describe the addition of flavour supermultiplets to a large $N$ theory with a holographic dual.  They are relatively easy to work with and admit simple dual descriptions and so are working examining.  There are four simple brane setups with an AdS$_3$ factor in the worldvolume of the probe brane; three of these have non-trivial Chern-Simons terms and the fourth does not.  As in~\cite{Davis:2008nv,Fujita:2009kw,HoyosBadajoz:2010ac}, the Chern-Simons terms arise from the Wess-Zumino (WZ) part of the brane action when the probe wraps a cycle with RR flux.

The simplest example is the 8 ND D3/D7 system.  This setup describes $(3+1)$-dimensional $\mathcal{N}=4$ $SU(N)$ SYM theory with$N_f$ $\mathcal{N}=(8,0)$ chiral flavours along a $(1+1)$-dimensional defect.  The holographic dual has $N_f$ D7 branes embedded in AdS$_5\times \mathbb{S}^5$.  When $N_f/N\ll 1$, the backreaction of the branes on the geometry can be neglected; this is the ``probe'' approximation.  In this case, the D7s wrap an AdS$_3$ cycle inside AdS$_5$ and the entire five-sphere.  Consequently, there is no transverse space in which the seven-branes can be separated from the stack of three-branes; the dual flavour is massless.  There is also a $U(N_f)$ gauge field $A$ on the D7 branes.  Reducing on the five-sphere, we find an infinite tower of fields in AdS$_3$.  One of these is the AdS$_3$ gauge field, which describes the $U(N_f)$ flavour symmetry currents of the dual theory.  The background metric can be written as
\begin{equation}
\label{ads5metric}
g=\frac{r^2}{l^2}(g_{1,1}+g_2)+\frac{l^2}{r^2}dr^2+l^2d\Omega_5^2, \hspace{.5in} l^4=4\pi g N \alpha'^2,
\end{equation}
where $g_{1,1}$ and $g_2$ denote the flat metrics on $\mathbb{R}^{1,1}$ and $\mathbb{R}^2$ respectively and $d\Omega_5^2$ is the metric on a unit five-sphere.  The probe branes wrap the five-sphere and the AdS$_3$ cycle formed by $\mathbb{R}^{1,1}$ together with the radial direction $r$.

When the $U(N_f)$ fields are small, they are well-described by the Yang-Mills action plus the WZ term,
\begin{equation}
S\approx -T_7(2\pi\alpha')^2\int d^8\xi \sqrt{-\text{P}[g]}\frac{\text{tr}(F_{mn}F^{mn})}{4}-\frac{T_7(2\pi\alpha')^2}{2}\int \text{P}[C_4]\wedge \text{tr}(F\wedge F),
\end{equation}
where $T_7$ is the tension of a D7 brane, $\xi^m$ labels the coordinates on the branes, $\text{P}$ indicates pullback to the brane worldvolume, $C_4$ is the background RR potential, and the trace is in the fundamental representation of $U(N_f)$.  Since the exterior derivative commutes with pullback, the WZ term can be integrated by parts to give
\begin{equation}
S_{WZ}=-\frac{T_7(2\pi\alpha')^2}{2}\int \omega_{3}[A]\wedge \text{P}[F_5],
\end{equation}
where
\begin{equation}
\omega_3[A]=\text{tr}\left(A\wedge dA+\frac{2}{3}A\wedge A\wedge A\right)
\end{equation}
is the Chern-Simons three-form and $F_5=dC_4$ is the RR five-form flux.  In order to find the effective action for the AdS$_3$ gauge field we must input the brane tensions and background fluxes.  With the conventions of~\cite{Davis:2008nv}, these are
\begin{align}
\label{probeConvention}
\int_{\mathbb{S}^5}F_5&=2\kappa_{10}^2 T_3N, \hspace{.9in} T_7=\frac{1}{g(2\pi)^3(2\pi\alpha')^4}, \\
2\kappa_{10}^2&=g^2(2\pi)^3(2\pi\alpha')^4, \hspace{.5in} T_3=\frac{1}{g(2\pi)(2\pi\alpha')^2}.
\end{align}
Using Eqs.~(\ref{ads5metric}) and~(\ref{probeConvention}) and the fact that the volume of the wrapped five-sphere is $\pi^3l^5$, the dimensionally reduced quadratic action for $A$ is then
\begin{equation}
S_A\approx - \frac{Nl}{32\pi}\int d^3x\sqrt{-G}\,\text{tr}F_{\mu\nu}F^{\mu\nu}-\frac{N}{4\pi}\int \omega_3[A],
\end{equation}
where $G$ is the metric on AdS$_3$.  As with the Kaluza-Klein gauge fields in the AdS$_3$ compactifications above, there is a propagating vector field of $m^2l^2=16$.  Per Eq.~(41) of Ref.~\cite{Larsen:1998xm}, it is dual to a primary operator of dimension $(3,2)$.  Meanwhile, the flat part of $A$ is dual to the $(1,0)$ $U(N_f)$ current algebra at level $N$.

The two other probe brane systems that feature Chern-Simons terms involve the addition of supersymmetric flavour to the ABJM theory~\cite{Aharony:2008ug}.  These systems were already studied in~\cite{Fujita:2009kw} and so I refer the reader there for more details.

There is one probe brane system with an AdS$_3$ factor where the three-dimensional action does \emph{not} have a Chern-Simons term.  This is the D3/D3 system~\cite{Constable:2002xt}.  I return to this in Section~\ref{chargedBTZ}

\section{Correlators from anomalies}
\label{anomalies}
The general results for $U(1)$ chiral currents at the end of Section~\ref{abelian} cry out for a purely field theoretic derivation.  Indeed, the fact that the current correlators were completely determined by the chiral anomalies suggests a good place to look.

In this section we will consider two-dimensional theories with a global $U(1)$ symmetry on an orientable Riemannian manifold with metric $g$.  A special fact about symmetries in two dimensions is that their symmetry currents may be related to (Hodge) dual symmetry currents.  This statement is beautifully illustrated in QED$_2$ with $N_f$ massless Dirac fermions.  Long ago, Schwinger solved thie theory \emph{exactly}~\cite{Schwinger}.  His result is essentially geometrical and consequently much more general. 

In two dimensions, the electric current $j^a=\bar{\psi}\gamma^a\psi$ is related to the axial current $j^a_A=\bar{\psi}\gamma^5\gamma^a\psi$ by $j^a_A=\epsilon^{ab}j_b$, since $\epsilon^{ab}\gamma_b=\gamma^5\gamma^a$ (with the orientation $\epsilon^{01}=1$.  In the language of differential geometry, the axial current is simply the Hodge dual of the electric current.  Moreover, the axial current is anomalous with
\begin{equation}
\partial_aj^a_A = \frac{N_f}{\pi}F_{01},
\end{equation}
where $F_{01}$ is the electric field strength and there are no other terms, since the axial symmetry is not explicitly broken.  On a curved background, a little more work shows that the axial current is still related to the electric current by Hodge duality, $j^a_A=\epsilon^{ab}j_b/\sqrt{g}$ (with $\epsilon^{01}=1$, as before).  Then in the language of differential forms, current conservation and chiral anomaly are simply expressed as
\begin{equation}
d\star j=0,\hspace{.3in} dj=-\frac{N_f}{\pi}F, \hspace{.3in} j=j_adx^a,
\end{equation}
where $\star$ is the Hodge star operator and $F=F_{01}dx^0\wedge dx^1$.

In fact this result is extremely general.  A non-anomalous $U(1)$ current is usually Hodge dual to an anomalous axial $U(1)$ current.  The axial symmetry, however, may also be broken explicitly.  For example, the electric symmetry in the two-dimensional Gross-Neveu and $\mathbb{CP}^{N}$ models is preserved but the axial symmetry is broken in the Gross-Neveu theory.  I will consider theories where the axial symmetry is only broken by the anomaly, whose strength is indexed by a positive integer $k$.  The conserved current satisfies two operator identities,
\begin{equation}
\label{wardId}d\star j=0,\hspace{.3in} dj=-\frac{k}{\pi}F,
\end{equation}
where I have normalized the second term so that it matches the normalization for the anomaly in Eq.~(\ref{gaugeInvJJ}).

Now consider the one-point function $J(x)=\langle j(x)\rangle$.  Eqn.~(\ref{wardId}) becomes a set of differential equations for $J$ that everyone knows how to solve.  If the manifold is simply connected, we introduce a potential by $\star J=d\phi$.  Then $\phi$ satisfies
\begin{equation}\label{phiEqn}
\star d\star d \phi = \star \frac{k}{\pi} F,
\end{equation}
For a general compact manifold, however, $\star J$ may not necessarily be exact.  In general
\begin{equation}
\star J = d\phi+\gamma,
\end{equation}
where $\gamma$ is a harmonic one-form.  For a genus $g$ Riemann surface, there are $2g$ such one-forms, corresponding to the $2g$ non-trivial one-cycles.  However, a short derivation shows that the harmonicity of $\gamma$ implies that both $d\gamma$ and $d\star \gamma$ vanish.  We thereby arrive at Eqn.~(\ref{phiEqn}) again for $\phi$.  In coordinates, this is just
\begin{equation}
-\partial_a(\sqrt{g}g^{ab}\partial_b\phi)=\frac{k}{\pi} F_{01}.
\end{equation}

Eqn.~(\ref{phiEqn}) can be solved by with a Green's function $G$ satisfying
\begin{equation}
-\partial_a(\sqrt{g}g^{ab}\partial_b)_xG(x,y)=\delta^{(2)}(x-y),
\end{equation}
to be
\begin{equation}
\phi(x)=\frac{k}{\pi}\int d^2y \,G(x,y)F_{01}(y)+c_0,
\end{equation}
where $c_0$ is a constant.  Then in coordinates the one-point functions of the vector current is
\begin{equation}
J^a(x)=-\frac{k\epsilon^{ab}}{\pi\sqrt{g}}\partial_b\int d^2y\, G(x,y)F_{01}(y)+ \gamma^a(x),
\end{equation}
where $\gamma$ is a harmonic one-form.  This result can be simplified further if our theory is generally covariant.  If so, we can employ a diffeomorphism to bring the metric into conformal gauge, $g_{ab}=e^{2\omega}\delta_{ab}$.  The covariant one-point function becomes
\begin{equation}
\label{covariantJ}
J_a(x)=-\frac{k}{\pi}\epsilon^{ab}\epsilon^{cd}\int d^2y \frac{\partial}{\partial x^b}\frac{\partial}{\partial y^d}G(x,y)A_c(y)+\gamma_a(x).
\end{equation}
The astute reader, recalling that $G$ can be expressed as the two-point function of a free massless scalar on the manifold, will note that Eq.~(\ref{covariantJ}) is the \emph{same} as that found in conformal field theory.  There a $U(1)$ current algebra is equivalent to a free massless scalar CFT: the current $j_a$ is obtained by differentiating the scalar, and so $J_a$ can be computed by linear response to be Eq.~(\ref{covariantJ}).

Eq.~(\ref{covariantJ}) precisely reproduces the non-perturbative results for current correlators but for an arbitrary manifold.  By calculating the exact one-point function of $J$ in the presence of arbitrary (symmetry-preserving) sources, we can compute \emph{all} connected correlators of $J$ with neutral operators\footnote{The correlators of $j$ with \emph{charged} operators are surely nonzero, but they cannot be computed with the method above.  To do so, we break the symmetry through the (small) source for the charged operator.}.  These are the correlators of $j$ with itself, the stress tensor, and those related by derivatives. These correlators are exactly those of a conformal theory.  Finally, note that the domain of validity for this computation (no charged sources) overlaps nicely with the gravity results of Section~\ref{abelian}.

As a final check of this result, let us work in flat space and expand in Fourier modes.  The Fourier space Green's function is just $1/q^2$ and so Eq.~(\ref{covariantJ}) becomes
\begin{equation}
J_a(q)=-\frac{k}{\pi}\frac{\epsilon^{ab}q_b\epsilon^{cd}q_d}{q^2} A_c(q)=\langle j_a(q) j^c(-q)\rangle A_c(q).
\end{equation}
This is precisely the result that follows from Eq.~(\ref{gaugeInvJJ}).

\section{External currents}
\label{chargedBTZ}
The Chern-Simons level for an AdS$_3$ gauge field must be an integer.  In the field theory this corresponds to the quantization of the level of the dual current algebra.  But what if the level vanishes?  A one-line argument using the current algebra Eq.~(\ref{kacMoody}) shows that such a theory is not unitary.  Usually we stop here, but there is an almost stupid loophole: what if the current is not an operator at all but rather a control parameter?  We could immediately enumerate a list of properties for such a theory.  First, it would correspond to a two-dimensional gauge theory to which we couple an \emph{external} current.  Consistency would require the source to be conserved.  The holographic dual would contain a gauge field with no Chern-Simons term, dual to the field theory gauge field.  The bulk action would contain the Maxwell term plus higher-derivative corrections.

But does such a system exist in string theory?  The answer is yes.  I now return to the D3/D3 setup.  The probe D3 branes describe the dynamics of $\mathcal{N}=(4,4)$ matter added to $\mathcal{N}=4$ SYM theory along a $(1+1)$-dimensional defect.  This system was originally studied in~\cite{Constable:2002xt} and examined for potential condensed matter applications in~\cite{Hung:2009qk}.  In the bulk, the probes wrap an AdS$_3$ cycle as well as an $\mathbb{S}^1$ inside the five-sphere.  The brane does \emph{not} wrap a cycle with RR flux and so the WZ part of the brane action does not give rise to a Chern-Simons coupling for the AdS$_3$ gauge field.

Indeed, the defect conformal theory on the D3/D3 intersection is fundamentally different from the other probe brane setups.  In the other probe systems, the probes had higher dimension than the colour branes and so in the decoupling limit (the $\alpha'\rightarrow 0$ limit when adding flavour to $\mathcal{N}=4$ SYM) the worldvolume fields on the probe brane along with gravity decouple from the theory on the colour branes.  With the 8 ND D3/D7 system of Section~\ref{8NDprobe}, the remaining dynamical fields arise from 3-3 and 3-7 strings.  The 3-3 strings describe the $\mathcal{N}=4$ SYM theory on the D3 branes and the 3-7 strings fundamental flavour.  In the D3/D3 intersection, however, the $\mathcal{N}=4$ SYM theories on both stacks of D3 branes remain dynamical in the decoupling limit.  The field theory is really $U(N_f)\times U(N)$ $\mathcal{N}=4$ SYM coupled to a bifundamental hyper along the defect.  Separating the two stacks in the transverse space does not give a mass to the bifundamentals, but rather corresponds to going out on the moduli space.  Finally, the 't Hooft coupling on the probe stack is $4\pi g N_f$, which goes to zero in the probe and supergravity limits.  The gauge theory on the probe stack then does not decouple \text{but} is arbitrarily weakly coupled.

This system is extremely interesting in its own right.  However, I want to draw attention to a simple fact: the theory on the D3/D3 intersection contains a $U(N_f)$ gauge field living on a $(1+1)$-dimensional defect.  This field is dual to the $U(N_f)$ gauge field on the probe D3s.  We have thus succeeded in finding a theory with a $(1+1)$-d gauge field to which we may couple an external current.  The gravitational dual behaves differently at a fundamental level than the Chern-Simons theories.

What are the physics of the gauge sector of theories like this one?  We begin to answer this question by studying the physics of the bulk theory.  To do so, we must first holographically renormalize it.  As we will see, this is a more subtle problem than usual.  Consequently there has been confusion on this point in the literature and so I will go into some detail.

\subsection{Holographic renormalization and a new Weyl anomaly}
Consider the Einstein-Maxwell action in $(2+1)$-dimensions with a metric $G$, and a $U(1)$ gauge field $A$,
\begin{equation}
\label{Seinmax}
S_{bulk}=\frac{1}{2\kappa^2}\int_M d^3x\sqrt{-G}(R-2\Lambda)+\frac{1}{\kappa^2}\int_{\partial M}d^2x\sqrt{-h}K-\frac{1}{4g^2}\int_M d^3x \sqrt{-G}F^2+S_{\rm CT}.
\end{equation} 
The bulk action, evaluated for a general solution to the equations of motion, will contain some divergences.  It is our task to add appropriate local and covariant counterterms on the AdS$_3$ boundary to cancel them as well as to yield a consistent variational principle.  Having done so, we may take well-defined variations of the bulk action with respect to boundary fields.

The equations of motion that follow from the action Eq.~(\ref{Seinmax}) are
\begin{align}
\label{eoms1}
R_{\mu\nu}-\frac{R}{2}G_{\mu\nu}+\Lambda G_{\mu\nu}&=\frac{\kappa^2}{g^2}\left( F_{\mu\rho}F_{\nu}^{\,\,\,\rho}-\frac{F^2}{4}G_{\mu\nu} \right), \\
\nonumber
D_{\mu}F^{\mu\nu}&=0.
\end{align}
In the presence of a cosmological constant $\Lambda = -1/l^2$, Eq.~(\ref{eoms1}) admits solutions of the form
\begin{align}
\label{seriesSol}
G&=\frac{r^2}{l^2}\left( g^{(0)}_{ab}(x)+\frac{g^{(2)}_{ab}(x)+h_{ab}(x)\ln r}{r^2}+\hdots \right)dx^adx^b+\frac{l^2}{r^2}dr^2, \\
\nonumber
A &= \left(\tilde{a}^{(0)}_a(x) \ln r + a^{(0)}_a(x)+\hdots\right) dx^a,
\end{align}
in a gauge with $A_r=0$ and the dots indicate terms suppressed by powers of $\ln r/r$ and $1/r$.  These are asymptotically AdS$_3$ geometries with a boundary at $r=\infty$.  There is a representative boundary metric $g^{(0)}$ of the induced conformal class on the boundary.  Moreover Marolf and Ross~\cite{Marolf:2006nd} have shown that this theory has only one possible quantization, namely the one where $\tilde{a}^{(0)}$ is the source.  It is easy to see that this is the only quantization consistent with conformal invariance.  We may demand that $a^{(0)}$ is free to fluctuate for constant $\tilde{a}^{(0)} \ln r$, but the converse depends on the choice of $r$ and so violates conformal invariance.

Before proceeding, let's employ the equations of motion to relate the parameters of the solution Eq.~(\ref{seriesSol}) to each other.  Plugging the series solution into the $rr$ component of Einstein's equations Eq.~(\ref{eoms1}) gives
\begin{equation}
h_{a}^a=0, \hspace{.5in} g^{(2)a}_{\,\,\,\,\,\,\,\,a}=-\frac{l^4}{2}R_2-\frac{\kappa^2l^2}{2g^2}(\tilde{a}^{(0)})^2,
\end{equation}
where $R_2$ is the Ricci scalar of the boundary metric $g^{(0)}$ which also contracts indices.  A similar analysis of the remaining equations of motion shows that, as usual, the indepedent boundary data are $g^{(0)}$, the trace-free part of $g^{(2)}$, $\tilde{a}^{(0)}$, and $a^{(0)}$.  However, this is not the whole story.  The $r$ component of Maxwell's equations shows that $\tilde{a}^{(0)}$ is conserved with respect to $g^{(0)}$, that is
\begin{equation}
D_a^{(0)}\tilde{a}^{(0)a}=0.
\end{equation}
where $D^{(0)}$ is the covariant derivative with respect to $g^{(0)}$.  This is important!  It tells us that $\tilde{a}^{(0)}$ is dual to the $U(1)$ current of the dual theory.  The subleading term $a^{(0)}$ would be the vev for the $U(1)$ gauge field in the boundary theory.  Moreover, since the only acceptable quantization is that where $\tilde{a}^{(0)}$ is fixed~\cite{Marolf:2006nd}, we see that the bulk theory precisely captures the fact that the dual theory has a gauge field in the presence of an \emph{external} source.

Now evaluate the bulk action Eq.~(\ref{Seinmax}) for the series solution Eq.~(\ref{seriesSol}), integrating up to a radial cutoff $r=r_{\Lambda}$.  The result is
\begin{equation}
\label{bulkDiverge}
S_{\rm bulk}=\int d^2x \sqrt{-g^{(0)}}\left[ \frac{r_{\Lambda}^2}{\kappa^2 l^3 }+\left(\frac{lR_2}{2\kappa^2}-\frac{(\tilde{a}^{(0)})^2}{2g^2l}\right)\ln r_{\Lambda} \right]+O(r_{\Lambda}^0).
\end{equation}
Before going on, take note of the logarithmic divergence.  This divergence is important: it implies a Weyl anomaly of the dual theory~\cite{Henningson:1998gx}.  Even after regulating all of the divergences, the variation of the bulk action under a infinitesimal Weyl transformation $G\rightarrow G(1+2\delta \omega)$, $r\rightarrow r(1-\delta \omega)$ is
\begin{equation}
\delta_W S_{\rm bulk}=\int d^2x\sqrt{-g^{(0)}}\delta\omega\left( -\frac{lR_2}{2\kappa^2}+\frac{(\tilde{a}^{(0)})^2}{2g^2l}\right),
\end{equation}
which implies a Weyl anomaly (as in~\cite{Henningson:1998gx}, in these conventions a free boson contributes to the anomaly as $-1/24\pi R_2$),
\begin{equation}
\label{weylAnom1}
T_a^a=\frac{l}{2\kappa^2}R_2-\frac{1}{2g^2l}(\tilde{a}^{(0)})^2.
\end{equation}
The first term gives the well-known central charge $c=12\pi l/\kappa^2$~\cite{Brown:1986nw} of the dual theory but the second is something new.  Actually, it was first observed but misinterpreted in~\cite{Jensen:2010vx}.  I'll return to this shortly.  For now let's finish the job of regulating the divergences in Eq.~(\ref{bulkDiverge}).  The first and second are easy to subtract: we simply add the boundary counterterms
\begin{equation}
S_{\rm CT,1}=-\frac{1}{\kappa^2l}\int_{r=r_{\Lambda}} d^2x \sqrt{-h}\left(1+\frac{R_hl^2}{2}\right),
\end{equation}
where $h$ is the induced metric on the cutoff slice and $R_h$ is the Ricci scalar formed from $h$.  The last divergence is a little subtle to regulate as, naively, there is no way to add a local and gauge-invariant boundary term to cancel it.
\subsubsection{Vector/scalar duality}
The simplest way to solve this problem is to employ vector/scalar duality.  The Einstein-Maxwell theory will become the theory of a minimally coupled massless scalar with gravity; we have known how to renormalize \emph{that} theory for a long time.  We can then translate those counterterms into those needed in the vector theory.

Let's Wick-rotate to the Euclidean theory.  Then the bulk field strength $F$ is related to the dual scalar $\phi$ by $F^{\mu\nu}=\epsilon^{\mu\nu\rho}\partial_{\rho}\phi/\sqrt{G}$ (with $\epsilon^{01r}=1$).  The gauge part of the action becomes
\begin{equation}
\label{Sphi}
S_{\phi}=-\frac{1}{2g^2}\int d^3x\sqrt{G}(\partial\phi)^2,
\end{equation}
so that $\phi$ has a near-boundary series solution
\begin{equation}
\phi(x,r)=\phi^{(0)}(x)+\frac{\phi^{(2)}(x)+\frac{l^4}{2}\Delta^{(0)}\phi^{(0)}(x)\ln r}{r^2}+...,
\end{equation}
where $\Delta^{(0)}$ is the scalar Laplacian with respect to $g^{(0)}$.  The functions $\phi^{(0)}$ and $\phi^{(2)}$ specify boundary data for $\phi$.  The on-shell bulk action Eq.~(\ref{Sphi}) is logarithmically divergent when $\phi^{(0)}$ has a gradient and is holographically renormalized with the counterterm
\begin{equation}
\label{phiCT}
S_{\text{CT},\phi}=\frac{l}{2g^2}\int_{r=r_{\Lambda}} d^2x\sqrt{h} (\partial \phi)_h^2\ln r_{\Lambda},
\end{equation}
where $(\partial\phi)_h^2$ involves a contraction with the induced metric $h$.  The logarithm again indicates that there is a Weyl anomaly in the dual theory, i.e.
\begin{equation}
\label{weylAnom2}
T_a^a=\frac{l}{2\kappa^2}R_2-\frac{l}{2g^2}(\partial \phi^{(0)})^2.
\end{equation}
This should not be surprising.  The data of the series solution for $A$, Eq.~(\ref{seriesSol}) are related to those for $\phi$ by
\begin{equation}
\label{bdyRelate}
\tilde{a}^{(0)a}=\frac{l\epsilon^{ab}\partial_b \phi^{(0)}}{\sqrt{g^{(0)}}}, \hspace{.5in}F_{ab}^{(0)}\equiv\partial_a a^{(0)}_b-\partial_b a^{(0)}_a = -\frac{2\epsilon_{ab}\phi^{(2)}}{l^3\sqrt{g^{(0)}}},
\end{equation}
where $\epsilon^{01}=1$ and indices are raised and lowered with $g^{(0)}$.  Then $(\partial \phi^{(0)})^2=(\tilde{a}^{(0)})^2/l^2$ and Eqs.~(\ref{weylAnom1}) and~(\ref{weylAnom2}) are equivalent.  

\subsubsection{Deriving the Weyl anomaly}
Can we understand the origin of the new Weyl anomaly in the scalar picture?  Yes; following~\cite{Petkou:1999fv} I will present a field theoretic derivation of the second term in Eq.~(\ref{weylAnom2}).  Recall that the massless pseudoscalar is dual to a marginal operator in the dual CFT.  The source for the dual operator is just $\phi^{(0)}$, which by Eq.~(\ref{bdyRelate}) is related to $\tilde{a}^{(0)}$ by $\tilde{a}^{(0)}=l\star d\phi^{(0)}$.  Earlier I noted that $\tilde{a}^{(0)}$ is proportional to the current in the boundary theory, so just as in Section~\ref{anomalies}, $\phi^{(0)}$ is a potential for the current.  The dual operator to $\phi$, $\mathcal{O}_{\phi}$, is then proportional to the electric field $F_{01}$.  Accordingly, a brief computation shows that the two-point function of $\mathcal{O}_{\phi}$  in the flat space vacuum is
\begin{equation}
\frac{\delta^2 S_{\rm ren}[\phi]}{\delta\phi^{(0)}(x)\delta \phi^{(0)}(0)}=\langle \mathcal{O}_{\phi}(x)\mathcal{O}_{\phi}(0)\rangle = \frac{2l}{\pi g^2 x^4},
\end{equation}
where I have defined $S_{\rm ren}[\phi]=\lim_{r_{\Lambda}\rightarrow\infty}\left( S_{\phi}+S_{\text{CT},\phi}\right)$.  

The astute reader may recall that the distribution $1/x^4$ is ill-defined in two dimensions.  It must be supplemented with a logarithmically divergent contact term in order to yield a well-defined Fourier transform, i.e.
\begin{equation}
\frac{1}{x^4}\rightarrow \frac{1}{x^4}+\frac{\pi}{4}\ln(x^2M^2)\Delta\delta^{(2)}(x),
\end{equation}
where $M$ is the renormalization scale and $\Delta$ is the flat space Laplacian.  The scale-dependence of the contact term gives rise to the Weyl anomaly.  To see this, consider the Callan-Symanzik Eq. (5) of Ref.~\cite{Petkou:1999fv} which in flat space reads
\begin{equation}
\sum_k \frac{1}{k!}\int d^2x_1\hdots d^2x_k \phi^{(0)}(x_1)\hdots \phi^{(0)}(x_k)M\partial_M\langle \mathcal{O}_{\phi}(x_1)\hdots \mathcal{O}_{\phi}(x_k)\rangle = \int d^2x \langle T_a^a(x)\rangle,
\end{equation}
There is an $M$-dependent contribution to the quadratic term in the LHS, which after integrating by parts gives 
\begin{equation}
T_a^a=-\frac{l}{2g^2}(\partial \phi^{(0)})^2,
\end{equation}  
precisely reproducing the relevant part of $T_a^a$ computed in Eq.~(\ref{weylAnom2}).  Indeed, Skenderis and Petkou worked out the general match between field theory and AdS/CFT for these ``contact'' Weyl anomalies in~\cite{Petkou:1999fv}.  The reader is encouraged to look there for more information.

\subsubsection{Wrapping up the vector theory}
Let's complete this discussion by finishing the renormalization of the vector theory.  To do this we simply translate the counterterm for the scalar Eq.~(\ref{phiCT}) in terms of $F_{\mu\nu}$.  The right counterterm is just
\begin{equation}
\label{counterterm}
S_{\text{CT},F}=\frac{1}{2g^2l}\int_{r=r_{\Lambda}} d^2x\sqrt{-h} F_{ra}F^{ra}\ln r_{\Lambda}.
\end{equation}
Such a counterterm, which in the gauge $A_r=0$ involves radial derivatives of the $A_a$, appears to be non-local on the boundary slice.  However it is manifestly local in the theory with the dual scalar and so must be local here as well.   The renormalized bulk action is then
\begin{equation}
\label{Sren}
S_{\rm ren}=\lim_{r_{\Lambda}\rightarrow\infty}\left( S_{\rm bulk}+S_{\rm CT,1}+S_{\text{CT},F}\right),
\end{equation}
from which we may take a variation $A_a\rightarrow A_a+\delta A_a$.  This variation may be decomposed into $\delta A_a= \delta\tilde{a}_a\ln r + \delta a_a + ...$, which leads to
\begin{align}
\delta S_{\rm ren}&=-\frac{1}{g^2l}\lim_{r_{\Lambda}\rightarrow\infty}\int_{r=r_{\Lambda}} d^2x\sqrt{-g^{(0)}}\left[\left(\delta \tilde{a}_a\ln r_{\Lambda}+\delta a_a\right) \tilde{a}^{(0)a}-\delta\tilde{a}_a \tilde{a}^{(0)a}\ln r_{\Lambda}\right], \\
\nonumber
&= -\frac{1}{g^2l}\int d^2x\sqrt{-g^{(0)}}\delta a_a\tilde{a}^{(0)a}.
\end{align}
The renormalized theory therefore corresponds to the grand canonical ensemble of the dual theory, since the relevant boundary condition is that $a^{(0)}_a$ is fixed.  This does not contradict the statement of Marolf and Ross~\cite{Marolf:2006nd} that the only quantization is the one where $\tilde{a}^{(0)}$ is an external source -- we are simply in a different ensemble.  Now I choose the normalization where I identify $a^{(0)}_a$ with the dual gauge field\footnote{The reader may note that $a^{(0)}_a$ transforms under bulk gauge transformations that have support at the boundary.  Such transformations are enacted by symmetry operators in the dual theory~\cite{Strominger:1997eq}.  The field $a^{(0)}$ therefore transforms in precisely the same way as the boundary gauge field and so we identify the two.} $\mathcal{A}_a$.  The one-point function of the current is
\begin{equation}
\label{jVEV}
\langle j^a\rangle = \frac{1}{\sqrt{-g^{(0)}}}\frac{\delta S_{\rm ren}}{\delta a_a}=-\frac{1}{g^2l}\tilde{a}^{(0)a}.
\end{equation}

We may also Legendre transform to a theory where the current is held fixed by adding a boundary term to the action.  That is, define a new bulk action by
\begin{equation}
\label{SrenLT}
\tilde{S}_{\rm ren}=\lim_{r_{\Lambda}\rightarrow\infty}\left( S_{\rm bulk}+\frac{1}{g^2l}\int_{r=r_{\Lambda}} d^2x \sqrt{-g}A_a F^{ra}+S_{\rm CT,1}-S_{\text{CT},F}\right),
\end{equation}
where the sign of the counterterm for the gauge field must also be flipped.  This boundary term simply adds the boundary density $-\sqrt{-g^{(0)}}a^{(0)}_a j^{a}$ and so is indeed just a Legendre transform of $S_{\rm ren}$.  The variation of the renormalized action under $\delta A_a=\delta\tilde{a}_a\ln r+\delta a_a+\hdots$ is then
\begin{equation}
\delta \tilde{S}_{\rm ren}=\frac{1}{g^2l}\int d^2x\sqrt{-g^{(0)}} \delta \tilde{a}^a a^{(0)}_a=-\int d^2x\sqrt{-g^{(0)}}\delta j^aa^{(0)}_a.
\end{equation}
where I have used Eq.~(\ref{jVEV}) to identify $\delta j^a=-\delta \tilde{a}^a/g^2l$.  This gives the one-point function
\begin{equation}
\label{aVEV}
\langle \mathcal{A}_a\rangle =-\frac{1}{\sqrt{-g^{(0)}}}\frac{\delta \tilde{S}_{\rm ren}}{\delta j^a}=a^{(0)}_a,
\end{equation}
as expected.  The action $\tilde{S}_{\rm ren}$ indeed corresponds to the canonical ensemble.  Moreover, the fact that $j^a$ is conserved means that the variation is only defined up to a total derivative.  That is, $a^{(0)}_a$ and $a^{(0)}_a-\partial_a \lambda$ are identified; $\mathcal{A}$ is in fact a gauge field.  Finally, note that the boundary term changes the Weyl variation of the bulk action.  We now have
\begin{equation}
\label{weylAnom3}
\delta_W \tilde{S}_{\rm ren}=-\int d^2x\sqrt{-g^{(0)}}\delta \omega \left( \frac{lR_2}{2\kappa^2}+\frac{(\tilde{a}^{(0)})^2}{2g^2l} \right)\rightarrow \tilde{T}_a^a=\frac{l}{2\kappa^2}R_2+\frac{g^2l}{2}j^2,
\end{equation}
so the new ``contact'' anomaly flips sign.

Let me make a final note of comparison with other works.  Rather than the counterterm Eq.~(\ref{counterterm}), three recent works~\cite{Hung:2009qk,Jensen:2010vx,Ren:2010ha} have employed the counterterm $S_{\rm CT,alt}=\sqrt{-h}A^2/\ln r_{\Lambda}$.  It turns out that this regularization is equivalent to the correct renormalization in the CE for the gauge $A_r=0$.  

\subsection{Application: two-point functions}
Now that we have developed this machinery let us use it.  First, I will compute the two-point function of the gauge field in the vacuum.  This correlator usefully characterizes the response of the plasma to an external source.  However it is counterintuitive: an electric field responds to the external current.  We may formally define a conductivity by linear response.

I will use the method of gauge-invariants~\cite{Kovtun:2005ev} to compute the two-point function.  This computation was done at in~\cite{Ren:2010ha}.  Here I clarify some details and lay out the problem in more generality for the benefit of future studies.  First, I elect to choose Lorentz gauge $q^a\langle \mathcal{A}_a(q)\hdots \rangle=0$.  The two-point function of the gauge field is then related to a single scalar function $G_F$ by
\begin{equation}
C^{\mathcal{A}}_{ab}(\omega,q)\equiv\langle \mathcal{A}_a(q)\mathcal{A}_b(-q)\rangle =\frac{\epsilon_{ac}q^c\epsilon_{bd}q^d}{q^4} G_F(q^2).
\end{equation}
$G_F$ is the two-point function of the electric field $F_{01}(q)=\epsilon^{ab}iq_a\mathcal{A}_b(q)$.  Before going on, note that in the Euclidean vacuum $G_F$ is the two-point function of a dimension $2$ operator and so is proportional to $q^2\ln q$.  The UV divergence is physical: it gives rise to the Weyl anomaly of the previous section.  Consequently $\langle \mathcal{A}_a \mathcal{A}_b\rangle$ has a logarithmic UV divergence.  This is similar to the logarithmic divergence of the current-current correlator in 4d, which corresponds to the Weyl anomaly proportional to $F^2$.

Now on to the general computation.  Consider the fluctuations of the bulk gauge field in the asymptotically AdS$_3$ metric
\begin{equation}
G=-r^2f(r)dr^2+r^2dx^2+\frac{dr^2}{r^2f(r)}, \hspace{.3in} A=0
\end{equation}
where I have set $l=1$ and $f(\infty)=1$.  The boundary action for $A$ in the CE Eq.~(\ref{SrenLT}) may be represented in terms of the gauge-invariant combination $F_{01}(\omega,q,r)=F_{01}$ via Maxwell's equations and the Bianchi identity to be
\begin{equation}
S_{\rm bdy}=\frac{1}{2g^2}\int_{r=r_{\Lambda}}\frac{d\omega dq}{(2\pi)^2}\frac{\sqrt{-h}f}{\omega^2-q^2f}\left[ \frac{F_{01}(F_{01}')^*}{r}-|F_{01}'|^2\ln r_{\Lambda}\right],
\end{equation}
where $F_{01}$ has the near-boundary form
\begin{equation}
\label{nearbdyF}
F_{01}(\omega,q,r)=\tilde{F}^{(0)}(\omega,q)\ln r + F^{(0)}(\omega,q)+\hdots
\end{equation}
and is a solution to Maxwell's equations.  An easy exercise shows that $F_{01}$ obeys the equation of motion
\begin{equation}
\label{maxwellF01}
F_{01}''+\left(\frac{1}{r}+\frac{f'}{f}\frac{\omega^2}{\omega^2-q^2f}\right)F_{01}'+\frac{\omega^2-q^2f}{f^2r^4}F_{01}=0.
\end{equation}
Moreover, the two-point function of $\mathcal{A}$ may be represented in terms of the derivatives of $S_{\rm bdy}$ with respect to $\tilde{F}^{(0)}(\omega,q)=i\epsilon^{ab}q_a\tilde{a}^{(0)}_b(\omega,q)$ by
\begin{equation}
C^{\mathcal{A}}_{ab}(\omega,q)=\epsilon_{ac}q^c\epsilon_{bd}q^d\frac{\delta^2 S_{\rm bdy}}{\delta \tilde{F}^{(0)}(\omega,q)\delta \tilde{F}^{(0)}(-\omega,-q)}.
\end{equation}
so that the second variation with respect to $\tilde{F}^{(0)}$ is just $G_F/q^4$.  A simple computation shows that $C^{\mathcal{A}}_{ab}$ is
\begin{equation}
\label{secondVarF}
C^{\mathcal{A}}_{ab}(\omega,q)=\frac{1}{g^2}\frac{\epsilon_{ac}q^c\epsilon_{bd}q^d}{\omega^2-q^2}\frac{\delta F^{(0)}(\omega,q)}{\delta \tilde{F}^{(0)}(\omega,q)}.
\end{equation}
The formal conductivity at $q=0$, $j_1(\omega)/F_{01}(\omega)$, is then $\sigma(\omega)=\frac{ig^2}{\omega}\frac{\delta \tilde{F}^{(0)}}{\delta F^{(0)}}.$

Eq.~(\ref{maxwellF01}) is exactly soluble in the AdS$_3$ vacuum and the BTZ black hole.  In the general case we impose retarded boundary conditions, i.e that $F_{01}$ is infalling at the horizon~\cite{Son:2002sd}.  For the vacuum we have $f=1$ and so
\begin{align}
\label{vacSol}
 F_{01}= c\left(J_0\left(\sqrt{\omega^2-q^2}/r \right)+i\text{ sgn}(\omega) Y_0\left(\sqrt{\omega^2-q^2}/r \right)\right),
\end{align}
which by Eqs.~(\ref{nearbdyF}) and~(\ref{secondVarF}) gives
\begin{align}
C^{\mathcal{A}}_{ab}(\omega,q)= \frac{\epsilon_{ac}q^c\epsilon_{bd}q^d}{\omega^2-q^2}\frac{-2\gamma+i\text{ sgn}(\omega)\pi-\ln\frac{\omega^2-q^2}{4}}{2g^2},
\end{align}
where we choose the $\text{sgn}(\omega)$ root of $\sqrt{\omega^2-q^2}$ when $q^2>\omega^2$.

\subsection{Application: the charged BTZ black hole}
There is another simple application that will clear up some more confusion in the literature.  Let us holographically renormalize the charged BTZ black hole and see what we learn from its thermodynamics.  This geometry is an asymptotically AdS$_3$ solution to the Einstein-Maxwell equations of motion Eq.~(\ref{eoms1}) and takes the form~\cite{Banados:1992wn}
\begin{align}
G&=-f(r)dt^2+r^2d\phi^2+\frac{dr^2}{f(r)}, \hspace{.5in} f(r)=r^2-r_h^2-\frac{g^2\kappa^2}{2}q^2\ln\frac{r^2}{r_h^2}, \\
\nonumber
A&=g^2q \ln \frac{r}{r_h}dt,
\end{align}
where I have set $l=1$ and $\phi\in [0,2\pi)$.  This is a black brane geometry with an outer horizon at $r=r_h$.  The black brane has a Hawking temperature and Bekenstein-Hawking entropy density of
\begin{equation}
\label{chargedBTZhawking}
T=\frac{2r_h^2-g^2\kappa^2q^2}{4\pi r_h}, \hspace{.5in} s= \frac{4\pi r_h}{2\kappa^2}.
\end{equation}
The black hole is extremal with nonzero entropy for $2r_h^2=g^2\kappa^2q^2$ -- in this limit the near-horizon geometry is AdS$_2\times \mathbb{R}$, as for higher-dimensional extremal AdS-Reissner-Nordstr\"om black holes.  It has been well known that the naive free energy of the black brane is logarithmically divergent~\cite{Martinez:1999qi}, a fact that has led to some confusion since the charged BTZ black hole was found.  We can now interpret and renormalize the divergence: it is simply due to the Weyl anomaly of the boundary CFT!

The gauge field has the near-boundary series expansion $A=g^2q(\ln r -\ln r_h)dt$, and so by Eqs.~(\ref{jVEV}),(\ref{aVEV}) we identify both the external current and gauge field in the boundary theory to be
\begin{equation}
\label{chargedBTZq}
j^0=q, \hspace{.5in} \mu\equiv\langle  \mathcal{A}_0\rangle = -g^2q\ln r_h.
\end{equation}
By also employing Eq.~(\ref{chargedBTZhawking}), we may write the free energy density in the CE, $\tilde{F}=-\tilde{S}_{\rm ren}$, of the charged BTZ black hole entirely in terms of field theory quantities
\begin{equation}
\tilde{F}=-\frac{1}{2}Ts+\frac{1}{2}\mu q+\frac{g^2}{4}q^2.
\end{equation}
Despite its strange form, this free energy is thermodynamically consistent.  To see this, employ Eqs.~(\ref{chargedBTZhawking}), and~(\ref{chargedBTZq}) to differentiate with respect to $-T$ and $q$ to obtain
\begin{equation}
-\frac{\partial \tilde{F}}{\partial T}=\frac{2\pi r_h}{\kappa^2}=s, \hspace{.5in}\frac{\partial \tilde{F}}{\partial q}=-g^2q\ln r_h=\mu,
\end{equation}
as they should be.  Moreover, the specific heat, $c_V=-T(\partial^2\tilde{F}/\partial T^2)_q$ is linear in $T$ for small $T$, just as the ordinary BTZ black hole.

To conclude this subsection I will say a few words about the microscopic counting of the black hole entropy in the semiclassical limit.  This was done in an existing claim~\cite{Cadoni:2007ck} some time ago.  However, the argument there misattributes the effects of the Weyl anomaly to a Casimir energy instead.  I will now show that their derivation may be adapted to account for this fact.  To do this I will employ a generalization of Cardy's formula to compute the density of CFT states with the black hole's mass and charge as in~\cite{Strominger:1997eq}.  This is the exponentiated entropy in the microcanonical ensemble.

The first step is to recognize that the mass of the charged BTZ black hole in the GCE, $\tilde{M}=\int d\phi(\tilde{F}+Ts)$, may be written as
\begin{equation}
\label{BTZmass}
\tilde{M}=\frac{3S^2}{4\pi^2c}+\frac{1}{2} \mu Q,
\end{equation}
where $S=2\pi s$ is the total entropy, $Q=2\pi q$ the total charge, and $c=12\pi/\kappa^2$ is the central charge of the dual CFT.  Notably the mass of the black hole in the GCE is just $M=\frac{3S^2}{4\pi^2c} -\frac{1}{2}\mu Q$.  This fact tells us something important: the Weyl anomaly in the measure of the path integral contributes precisely $\frac{1}{2}\mu Q$ to the black hole mass.  It seems reasonable then that the charged BTZ black hole corresponds to a field theory state with $L_0$ and $\bar{L}_0$ eigenvalues that are related just to the first term of Eq.~(\ref{BTZmass}).

Recall that a state of a 2d CFT with eigenvalues $L_0$ and $\bar{L}_0$, in the normalization where the $M=Q=0$ vacuum (i.e. global AdS$_3$~\cite{Strominger:1997eq}) has $L_0=\bar{L}_0=0$, has a mass $M=L_0+\bar{L}_0$ and spin $J=L_0-\bar{L}_0$.  The charged BTZ black hole then represents a state with
\begin{equation}
\label{L0btz}
L_0=\bar{L_0}=\frac{3S^2}{8\pi^2c}.
\end{equation}

Next, we recall that the Hamiltonian in the CE is just that of the original theory subject to the constraint that the charge is fixed.  The dual theory does not have a gravitational anomaly and so should be modular invariant, provided that the current transforms accordingly.  The partition function of the theory at nonzero charge should then be related to a high temperature limit subject to a nonzero current.  We may then count the asymptotic growth of states for $S\gg c$ directly by the usual Cardy formula~\cite{Cardy:1986ie},
\begin{equation}
\rho(L_0,\bar{L}_0)\sim \exp\left[ 2\pi\sqrt{\frac{c L_0}{6}}+2\pi\sqrt{\frac{c\bar{L}_0}{6}}\right],
\end{equation}
which indeed gives by Eq.~(\ref{L0btz})
\begin{equation}
\rho\sim e^{S},
\end{equation}
``reproducing'' the black hole entropy.

This calculation has a few deficiencies.  The first is the calculation of $L_0$ and $\bar{L}_0$ for the black hole.  It seems reasonable to separate the mass into a piece that comes from the Weyl anomaly and a piece that is determined by the $L_0$ and $\bar{L}_0$ eigenvalues of the state, but it would be nice to verify it directly.  Second and more importantly, what is the correct modular invariant partition function of this theory?

\section{Conclusions}
The chiral anomaly for symmetry currents in 2d CFTs has some remarkable implications.  The correlators of $U(1)$ currents Eqs.~(\ref{gaugeInvJJ}),~(\ref{covariantJ}) are completely determined by the anomaly as long as the symmetries are broken only by the anomaly.  As a consequence the $U(1)$ currents are essentially non-interacting; their modes only lie on the lightfront and so the currents never behave hydrodynamically.  In a holographic dual, this fact is captured by the Chern-Simons term that encodes the anomaly.  In the field theory it is a geometric result.

Moreover, as a consequence of the bulk Chern-Simons term, charged black holes look very different in three dimensions compared to their higher-dimensional cousins.  They are simply black holes supplemented with a flat connection.  Taken together these facts have some implications for the AdS/CMT program in two dimensions (specifically applied for 2d CFTs with current algebras).  First, there is no naive superfluid instability: the fluctuation spectrum of the bulk theory does not see any instabilities when shifting the chemical potential, $\mu$ because this just a flat shift of the spatial component of the dual gauge field.  There may, however, exist thermodynamically preferred states with charged condensates at large enough $\mu$ -- it will take numerical study to answer this question.  Similarly, there should be no ``non-Fermi-liquid'' behavior at nonzero density~\cite{Cubrovic:2009ye,Liu:2009dm} -- the spectral functions of charged fermions are only shifted in spatial momentum by flat shifts of $A_x$\footnote{This problem was studied in a new preprint~\cite{Balasubramanian:2010sc}.}.  Second, in order for even the \emph{possibility} of non-trivial physics with $U(1)$ currents to exist we need to radically deform the bulk theory.  One possibility is to break Lorentz-invariance in the UV, breaking the chiral anomaly as well.  The gravitational dual would have non-AdS$_3$ asymptotics near the boundary~\cite{Son:2008ye,Balasubramanian:2008dm,Kachru:2008yh}.  For parity-invariant theories with a $U(1)_V\times U(1)_A$ symmetry like Eq.~(\ref{paritySeff}) in the D1/D5 system, we may also explicitly break the axial symmetry.  Presumably this would be dual to the explicit breaking of the bulk gauge symmetry corresponding to the $U(1)_A$ symmetry.

Even with the dynamics of the currents essentially fixed by the chiral anomaly, there are still a few non-trivial axes in the phase diagram of these theories.  One such axis involves the higher dimension vector primary that accompanies the current algebra.  Recall that this operator is dual to the propagating mode of the bulk gauge field.  For gravitational duals where the dual operator is relevant we may turn on a source for it.  Indeed, the case where the anomaly vanishes that I considered in Section~\ref{chargedBTZ} may be thought of as such a system in the limit that the vector primary becomes a dimension $1$ primary, i.e. the gauge field.  Another is a double-trace deformation $\mathcal{O}^{\dagger}\mathcal{O}$ for a charged operator $\mathcal{O}$ of appropriately low dimension~\cite{Witten:2001ua}.  A negative double-trace typically triggers an instability in the fluctuation spectrum of the dual operator -- it condenses.  Thus a holographic superfluid phase should be triggered by a suitable double-trace deformation~\cite{Faulkner:2010gj}.  Finally, we may add the marginal double-trace deformations for the current, $\text{tr}(j^2)$.  When the theory is parity-invariant we may add $\text{tr}(j_zj_{\bar{z}})$ which simply shifts the coupling $\lambda$ of the boundary WZW model Eq.~(\ref{chiralWZW}) away from the fixed point value.

The dynamics when the anomaly vanishes are more interesting.  Indeed, there are a few stringy AdS$_3$ compactifications that include gauge fields without Chern-Simons terms.  I discussed one of these examples, the D3/D3 intersection, in Section~\ref{chargedBTZ}.  Another example was found in a preprint~\cite{Colgain:2010rg} released just a few days ago in the context of wrapped M5 branes.  Since the anomaly for a symmetry current of a 2d CFT must be nonzero, the bulk gauge fields in these setups are dual to \emph{gauge} sectors to which we may couple an external current.  This claim was easy to verify for the D3/D3 intersection; it would be nice to do so directly for the dual field theory of~\cite{Colgain:2010rg}.

In Section~\ref{chargedBTZ} I took some steps toward a proper analysis of these systems.  First, I holographically renormalized the gravitational theory.  This machinery is required to precisely relate bulk quantities to observables in the boundary theory.  These results will therefore form a necessary stepping stone for the future study of these theories.  Next, I used these results to learn some physics.  In particular, there is a Weyl anomaly in the presence of an external current, Eq.~(\ref{weylAnom1}).  This anomaly is essentially due to an ultra-local divergence in the two-point function of the dimension two electric field, $\langle E(x) E(0)\rangle \sim 1/x^4$.  I concluded with some analysis of the charged BTZ black hole~\cite{Martinez:1999qi}.  The logarithmic divergences related to this geometry have confused people for some time.  We may now interpret them as simply arising from the Weyl anomaly of the boundary CFT.

\acknowledgments
I continue to be deeply indebted to Andreas Karch for his numerous insights.  I also thank Steve Carlip, Carlos Hoyos, Chris Herzog, Pavel Kovtun, Adam Ritz, and Dam Son for useful conversations and correspondence.  This work was supported in part by the U.S. Department of Energy under Grant Numbers DE-FG02-96ER40956 and DE-FG02-00ER41132 as well as NSERC, Canada.

\bibliography{2d}
\bibliographystyle{JHEP}

\end{document}